\documentclass[namedreferences,hyperref,optionalrh,solaromanenum]{spr-sola}
\usepackage{graphicx}	
\usepackage{amsmath}	
\usepackage{amsfonts}
\usepackage{multicol}   
\usepackage{bm}		    
\usepackage{breakurl}   

\newcommand{\hl}[1]{#1}

\newcommand{\BoldVec}[1]{\mathchoice%
  {\mbox{\boldmath $\displaystyle     #1$}}%
  {\mbox{\boldmath $\textstyle        #1$}}%
  {\mbox{\boldmath $\scriptstyle      #1$}}%
  {\mbox{\boldmath $\scriptscriptstyle#1$}}%
}
\newcommand{\EQ}{\begin{equation}}
\newcommand{\EN}{\end{equation}}
\newcommand{\EQA}{\begin{eqnarray}}
\newcommand{\ENA}{\end{eqnarray}}

\newcommand{\Fig}[1]{Fig.~\ref{#1}}

\newcommand{\xx}{\BoldVec{x}{}}

\newcommand{\BB}{\BoldVec{B} {}}

\newcommand{\AAA}{\BoldVec{A} {}}

\newcommand{\JJ}{\BoldVec{J} {}}
\newcommand{\nn}{\BoldVec{n} {}}
\newcommand{\ee}{\BoldVec{e} {}}

\newcommand{\nab}{\BoldVec{\nabla} {}}

\newcommand{\llll}{\BoldVec{l} {}}

\newcommand{\GGGG}{{\bf G} {}}
\newcommand{\dd}{{\rm d} {}}

\graphicspath{{./fig/}{./png/}}

\begin{document}

\begin{frontmatter}
\title{Magnetic vector potentials and helicity in periodic domains}

\author[addressref=aff,email={simon.candelaresi@gmail.com}]{\inits{S.C.}\fnm{Simon}~\snm{Candelaresi}\orcid{0000-0002-7666-8504}}
\address[id=aff]{High-Performance Scientific Computing,
University of Augsburg,
Universit\"atsstra\ss e 12a,
Augsburg,
86159,
Germany
}

\runningauthor{S. Candelaresi}
\runningtitle{Magnetic vector potentials and helicity in periodic domains}

\begin{abstract}
For a magnetic field in three-dimensional Euclidean space, we argue about
the existence of a magnetic vector potential in a finite cuboid
with periodic boundary conditions.
We first argue in physical terms by representing periodic domains as tori
and show that in some cases, the generating electric current
is unphysical.
To formalize our results we then represent magnetic fields and vector
potentials as differential
$2$ and $1$-forms in three-dimensional periodic domains.
\hl{
Unlike previous works we do not require the full formalism of the
de Rham cohomology for the doubly periodic case.
}
For a given magnetic field we ask the question whether a corresponding
vector potential exists.
Singly periodic domains are shown to always allow for a vector potential.
For doubly periodic domains it only exists if the net magnetic flux
through the non-periodic dimension vanishes, while for triply periodic domains
the net flux must vanish in all directions.
This has implications for the definition of magnetic helicity in periodic domains.
There we show that periodic domains together with a mean magnetic flux lead to contradicting results
about the value of the magnetic helicity and we show in a few thought experiments
fundamental contradictions to the interpretation as the linking of magnetic field lines.
For the helicity we can even construct magnetic fields in periodic domains
for which the magnetic helicity exists, but the calculations of the magnetic
helicity \hl{do} not lead to meaningful results due to the gauge freedom.
With that we conclude that in presence of a mean magnetic field the calculated magnetic
helicity \hl{cannot be interpreted geometrically in a straightforward way.
However, using the winding gauge we can recover the geometric interpretation
of the helicity as the flux averaged magnetic field line winding.
We showcase this further for a magnetic field with non-periodic
boundaries where the point-wise flux does not vanish while the surface
integrated flux vanishes.
This is relevant in the solar context.
Using an appropriate gauge transformation, the magnetic helicity
can be shown to vanish.
The winding helicity, however, would lead to a non-vanishing value here.
}
\end{abstract}


\end{frontmatter}

\section{Introduction}

Magnetic fields are used in a vast number of scientific areas.
They appear in particle physics, plasma physics, geology, astrophysics,
biology, atomic physics, fusion science, electric engineering and many more.
For modeling the dynamics of such physical systems it is more often than not necessary
to solve partial differential equations numerically.
The form of the equations depends on the subject considered and ranges from
simple mean-field equations \citep{steenbecketal66, Krause-Radler-1971-SolMagFields,
Leorat-Frisch-Pouquet-1975-AnnNYAcadSoc, Gruzinov1994}
over magnetohydrodynamics (MHD) \citep{Alfven-1942-150-405-Nature, Chandrasekhar1961hhs,
Frisch-Pouquet-Leorat-1975-JFluidMech, BrandenbSubramanianReview2005}
to relativistic \citep{Zrake-2014-794-2-ApJL} multicomponent fluids
with charged inertial particles \citep{Zakharov-Rogers-1992-4-10-PhysFluidB,
Suess-Wang-1999-104-A3-JGRA}.

One of the fundamental Maxwell-equations within MHD is the divergence-freeness of the
magnetic field $\BB$:
\EQ \label{eq: div B}
\nab\cdot\BB = 0,
\EN
which is equivalent to the non-existence of magnetic monopoles.
Since $\nab\cdot\nab\times\AAA = 0$ for any doubly differentiable field $\AAA$, we expect
to be able to find a magnetic vector potential $\AAA$ such that $\nab\times\AAA = \BB$.

The existence of such a field depends on the topology of the domain
and the presence of a mean magnetic flux.
As we will point out, it is restricted by the existence of periodic directions.
To show that we first argue in physical terms by representing the domains
as tori and the generating electric current being outside the domain.
Then we make use of differential forms and the Poincar\'e's lemma;
two concepts from differential geometry to confirm the results.

\hl{
For the doubly periodic case we arrive at the same result as
\cite{Pfefferle-Noakes-2021-62-9-JMthPhys} who applied
the full formalism of the de Rham cohomology in their Propositions
III.7 and III.8.
Moving to the triply periodic case, the existence of a vector potential
is a consequence of $T^3$ being cohomologous to $\mathbb{R}^3$
\citep{Pfefferle-Noakes-2021-62-9-JMthPhys}.
However, here we arrive at this result using only a simple
differential forms argument rather than the entire machinery
of cohomology.
}
We finalize our discussions by showcasing implications for the magnetic
helicity, a measure of the linkage of magnetic field lines.

The magnetic helicity $H_{\rm m} = \int_V \AAA\cdot\BB \ \dd V$
measures the degree of magnetic field line linking, knotting, braiding
and twisting \citep{MoffattBook1978}.
It is a measure for the field line topology.
If $\AAA$ is not well defined, then this quantity won't be either.
However, from its topological interpretation we can compute
the magnetic helicity through the
Gauss linking integral even without knowing $\AAA$.
This puts us into the dangerous position where we compute something
that might not exist, as was pointed out by
\cite{Berger-1996-34-225-ApLC} for a doubly-periodic
domain in three dimensions.
While \cite{Berger-1996-34-225-ApLC} made the argument that one could
invert the sign of the magnetic helicity through a series of suitable
magnetic reconnection events, he did not realize that the fundamental
issue of his test case is the non-existence of $\AAA$.

A discussion purely about the magnetic vector potential
would ignore
the possible scenarios where we can indeed compute $\AAA$, while the
calculation of the
\hl{
interpretation of the magnetic helicity as the Gauss linking number fails.
}
We show this through a series of simple examples at the end of our paper
\hl{
and highlight how in such scenarios the winding helicity
\citep{Prior-Yeates-2014-787-2-ApJ, Xiao-2025}
can be still applied with its interpretation as the flux averaged
magnetic flux tube winding.
}

\section{Physical Considerations}

A cuboid of dimension $n$ with $k \le n$ many periodic directions is topologically
equivalent to the domain $D^{n-k}\times T^k$, with the $(n-k)$-dimensional hyperrectangle
$D^{n-k}$ and the $k$-dimensional torus $T^k$.
This is easily seen for the case of a two-dimensional domain, which
can be represented as the cylinder $D\times S$, where $D = [a, b]$
with $a, b \in \mathbb{R}$ and $a < b$.

For a static magnetic field we can compute the generating electric current
density $\JJ = \nab\times\BB$ from the magnetic field $\BB$.
This method, however, breaks down in a finite domain in presence of a mean
magnetic field.
Take, for instance, a homogeneous field for which the curl operation vanishes.
The current from $\nab\times\BB$ is clearly $0$, which cannot be true for the
entire domain, since the current generates the magnetic field.
This problem is reflected when using Biot-Savart's law
\EQ \label{eq: biot savart}
\BB(\xx) = \frac{\mu_0}{4\pi} \int_{V} \frac{\JJ(\xx')\times(\xx-\xx')}{|\xx-\xx'|^3} \ \dd^3 x',
\EN
with the magnetic permeability of vacuum $\mu_0$.
Hence, a necessary condition for a mean-magnetic field to exist is the presence
of an electric current density $\JJ$ {\em outside} the domain.

Here we discuss three cases for three different sets of periodic boundary conditions.
The mean magnetic fields are then generated by external currents.
We discuss the nature of those currents \hl{and their physicality}.

\subsection{Singly Periodic Domains}

\begin{figure}[t!]\begin{center}
\includegraphics[width=0.5\columnwidth]{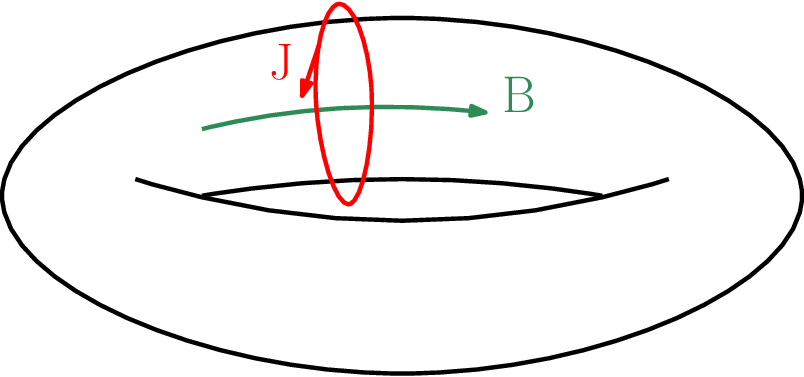}
\end{center}
\caption[]{
Representation of a singly periodic domain as solid torus.
The periodic dimension runs along the major axis.
A mean magnetic field in that direction can be generated
by an external poloidal electric current.
}
\label{fig: singly_periodic_as_torus}\end{figure}

For a three-dimensional cuboid with one periodic direction
we can find a representation as a solid torus
with the major axis being the periodic direction.
Any magnetic field within the box can then be generated by currents within
or by external currents.
External currents are needed in case of a mean magnetic field, i.e.\ net magnetic flux,
through one of the boundaries.
Such an external current can be a poloidal one, which generates
a mean magnetic field in the direction of the major axis (periodic direction).
Circular currents running along the major axis can generate mean magnetic fields
along the non-periodic axis (see \Fig{fig: singly_periodic_as_torus}).
For the singly periodic case we see that \hl{in any} case we can construct an external
current, which generates the magnetic field.

\subsection{Doubly Periodic Domains}

\begin{figure}[t!]\begin{center}
\includegraphics[width=0.95\columnwidth]{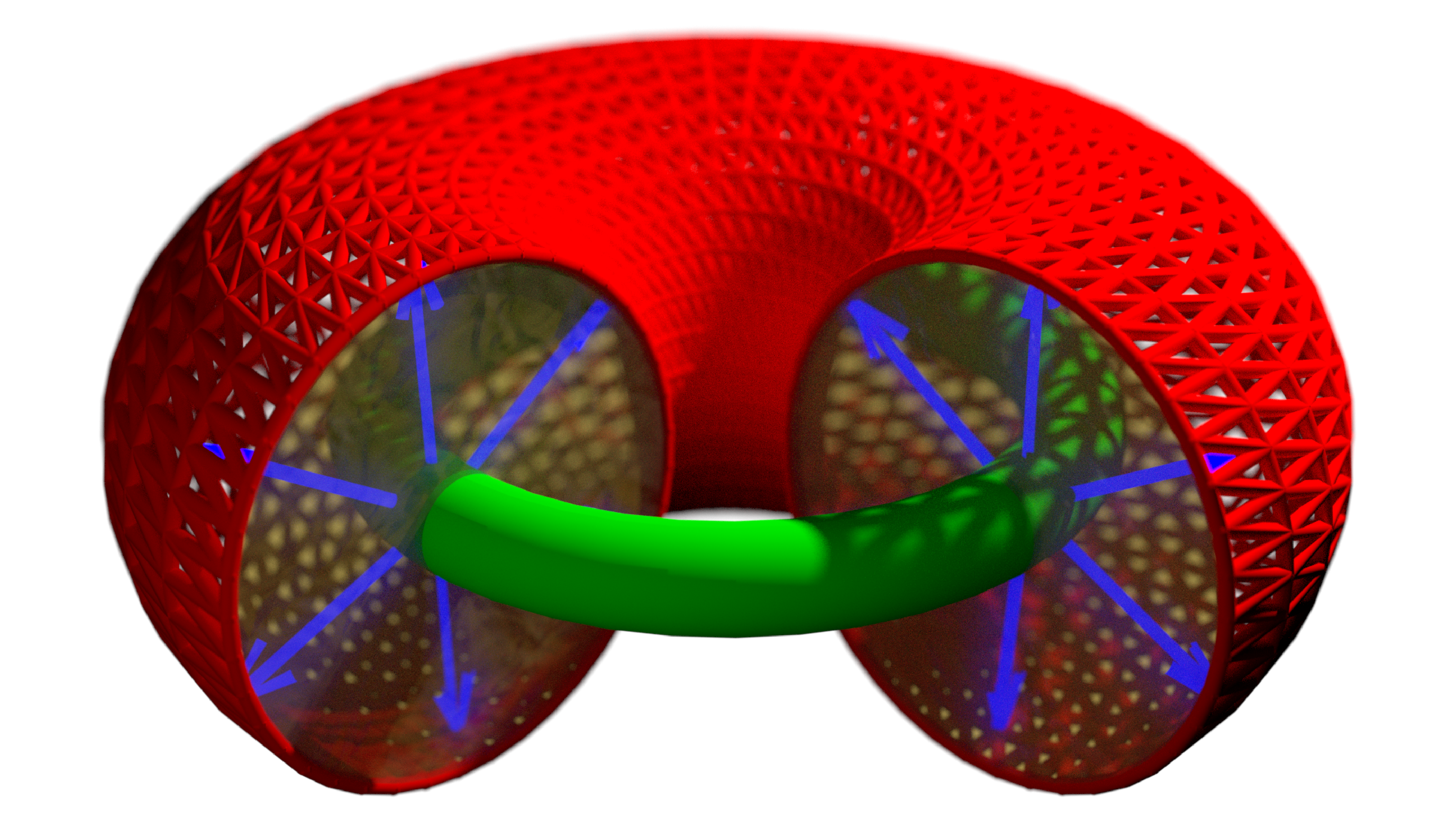}
\end{center}
\caption[]{
Representation of a doubly periodic domain as $[0,1] \times S \times S$.
The non-periodic dimension extends from the inner (green) to the outer (red)
torus.
With blue arrows we highlight a homogeneous magnetic field for which a magnetic
vector potential cannot be obtained.
}
\label{fig: torus}\end{figure}

For the doubly periodic domain we can construct a topologically equivalent domain
as $D \times S\times S = D \times T^2$, where $D = [a, b]$ is a real interval
with $a, b \in \mathbb{R}$ and $a < b$.
This domain can be represented as a solid torus with a hollow torus cut out
(see \Fig{fig: torus}).
Since we have two periodic directions, we could have chosen either periodic
direction as the major and minor axis of this torus.

Any mean magnetic field through one of the periodic directions
can then be constructed via an external electric current, similarly
to the singly periodic case.
A mean magnetic field through the non-periodic direction, however, is only
possible by placing a net magnetic monopole distribution at the center of the
hollow region, as was already pointed out by \cite{Berger-1996-34-225-ApLC}.

This can be also seen by computing the magnetic flux through a closed surface,
which does not vanish for all such surfaces.
One of those is the torus surrounding the inner torus of our domain.
In this embedding of the domain, this means that the magnetic field is not divergence free.
Hence, we have shown that a mean magnetic field in the non-periodic direction for
the doubly periodic domain is physically not meaningful.

\subsection{Triply Periodic Domains}

Triply periodic domains are topologically equivalent to the three-torus $T^3$.
Since $T^3$ needs to be embedded into $\mathbb{R}^4$ or higher, the above way of arguing
is not going to be very useful.
Instead, we consider the triply periodic domain as an infinite copy
of itself stacked onto all directions.

Suppose there was no mean magnetic field, i.e.\ the surface integral of $\BB$
on any closed or periodic surface vanishes.
The electric current $\JJ = \nab\times\BB$ inside the domain can then completely generate the magnetic
field using Biot-Savart's law.

With a net magnetic flux through any of the domain boundaries we can decompose the field into a component
with vanishing flux and a homogeneous component that gives rise to the net flux.
In our picture of the infinite copies of the domain
a homogeneous magnetic field translates into the entire space $\mathbb{R}^3$ being
filled with the field.
If we think of the field in each such periodic box being generated by a
current at the boundary of the box, we can construct the mean component of $\BB$ via $\JJ\propto\frac{I}{r}$
with the current $I$ and radius $r$.
As we take the limit $\lim_{r\rightarrow\infty}$ the necessary current diverges.
This of course is physically not meaningful.

\section{Magnetic Field and Vector Potential as Differential Forms}

In Euclidean geometry we can associate a $2$-form to the magnetic field $\BB(\xx)$ of
the form
\EQA
\beta(\xx) & = & B_1(\xx) \dd x_2 \wedge \dd x_3 + B_2(\xx) \dd x_3 \wedge \dd x_1 \nonumber \\
 & & + B_3(\xx) \dd x_1 \wedge \dd x_2,
\ENA
where $\dd$ is the exterior derivative and $x_{i}$ are the components of the
position vector $\xx$.
Taking the exterior derivative of $\beta$ results in
\begin{eqnarray}
& \dd\beta = &
(\partial B_1(\xx)/\partial x_1 + \partial B_2(\xx)/\partial x_2 +
\partial B_3(\xx)/\partial x_3) \nonumber \\
& & \times \dd x_1 \wedge \dd x_2 \wedge \dd x_3,
\end{eqnarray}
which is the $3$-form associated with the divergence of $\BB$.
From $\nab\cdot\BB = 0$ it follows that $\beta$ is a closed form, i.e.\ $\dd\beta = 0$,
a necessary condition for a potential to exist.

For $\beta$ the potential $\alpha$, with $\dd\alpha = \beta$, is
a $1$-form associated to the magnetic vector potential $\AAA(\xx)$ of
the form
\EQ
\alpha(\xx) = A_1(\xx) \dd x_1 + A_2(\xx) \dd x_2 + A_3(\xx) \dd x_3,
\EN
provided it exists.
Its differential
\begin{eqnarray}
\dd \alpha & = & (\partial A_2/\partial x_1 - \partial A_1/\partial x_2) \dd x_1 \wedge \dd x_2 + \nonumber \\
&& (\partial A_3/\partial x_2 - \partial A_2/\partial x_3) \dd x_2 \wedge \dd x_3 + \nonumber \\
&& (\partial A_1/\partial x_3 - \partial A_3/\partial x_1) \dd x_3 \wedge \dd x_1 \nonumber \\
& = & \beta
\end{eqnarray}
is equivalent to $\nab\times\AAA = \BB$.

Poincar\'e's Lemma states that if $\dd\beta = 0$, i.e.\ $\beta$ is closed,
on a contractible manifold then there exists a potential $\alpha$ such that
$\dd\alpha = \beta$, i.e.\ $\beta$ is exact.
Contractible means that every two-dimensional closed surface in the domain
can be smoothly mapped into a point lying in the domain without leaving
the domain at any stage of the transformation.
In other words, the Betti number of second kind vanishes, i.e.\ $b_2 = 0$.

Many practical domains are contractible, like the cuboid and solid ball.
Non-contractible domains that sometimes appear in applications are the spherical shell
and the solid torus.
Periodic boxes might first appear to be contractible, however, due to their
topology they can well be non-contractible, like the toroidal shell
for the doubly periodic domain.

Should the domain not be contractible, i.e.\ $b_2 \ne 0$ we can make use of a second sufficient
condition that states that if the integral of $\beta$ on all fundamental periods dimension $2$
vanishes then $\beta$ is exact.
For that we have to identify all fundamental periods of dimension $2$ of the domain.
In general terms, $2$-dimensional periods are all $2$-dimensional surfaces that lie within the domain
but cannot be contracted smoothly into one point without leaving the domain.
If one period can be smoothly transformed into another without leaving the domain they
are said to be linearly dependent.
The set of the linearly independent periods forms the set of the fundamental periods.
The number of fundamental periods of dimension $2$ is the Betti number of second kind $b_2$.
For instance, a cuboid with two holes has two fundamental periods, i.e.\ $b_2 = 2$.

Integrals on two-dimensional manifolds of $\beta$ of the form $\int_L \beta$ with the
manifold $L$ are the surface integrals for the normal component of the magnetic field
\EQ
\int_L \beta = \int_L \BB(\xx)\cdot\nn \ \dd A.
\EN
This equivalence between the integral of the two-form $\beta$
and the normal component of the magnetic field
makes the verification of the existence of potentials easier.

\subsection{Singly Periodic Domains}

Domains with one periodic dimension are topologically equivalent to a
solid torus $D^2 \times S^1$.
For such a domain, any two-dimensional closed surface is contractible
and the Betti number of second kind vanishes.
Consequently, for any closed $2$-form $\beta$ there exists a potential $\alpha$
such that $\dd\alpha = \beta$.
With that there exists a magnetic vector potential $\AAA$ defined in the whole
domain such that $\nab\times\AAA = \BB$.

\subsection{Doubly Periodic Domains}

Topologically, doubly periodic domains are equivalent to a solid torus with a hole
$[0, 1] \times S \times S$ (\Fig{fig: torus}) as was pointed out by
\cite{Berger-1996-34-225-ApLC}.
The two periodic directions run along the major and minor axis while the non-periodic
direction ranges from the inner to the outer torus.

To find the Betti number we make use of Poincar\'e polynomials.
For every topological space $X$ we can find the generating Poincar\'e polynomial
\EQ
P_X(z) = \sum_i b_{i}z^i,
\EN
where the coefficients $b_i$ are the Betti numbers of $i$th kind
and $z \in \mathbb{C}$.
Any direct sum of two topological spaces $X$ and $Y$ has the Poincar\'e
polynomial $P_{X\times Y} = P_{X}P_{Y}$.
For the circle $S$ we know that $P_S = 1 + z$ and for the line $P_{[0,1]} = 1$.
With that we can compute the Poincar\'e polynomial for the space
$[0,1] \times S\times S$ to
\EQ
P_{[0,1] \times S \times S} = 1 + 2z + z^2
\EN
and notice that the Betti number of second kind is $b_2 = 1$.
This is also seen from \Fig{fig: torus} where we can add a torus around
the inner cavity which is not contractible.
This torus is also the fundamental period for which the integral $\int_L \beta$
has to vanish for $\alpha$ to exist.

With only a net magnetic flux along either periodic dimension this integral indeed
vanishes.
For a net flux along the non-periodic direction, however, this integral does
not vanish.
Consequently, for such fields a magnetic vector potential does not exist.

\subsection{Triply Periodic Domains}

\begin{figure}[t!]\begin{center}
\includegraphics[width=0.7\columnwidth]{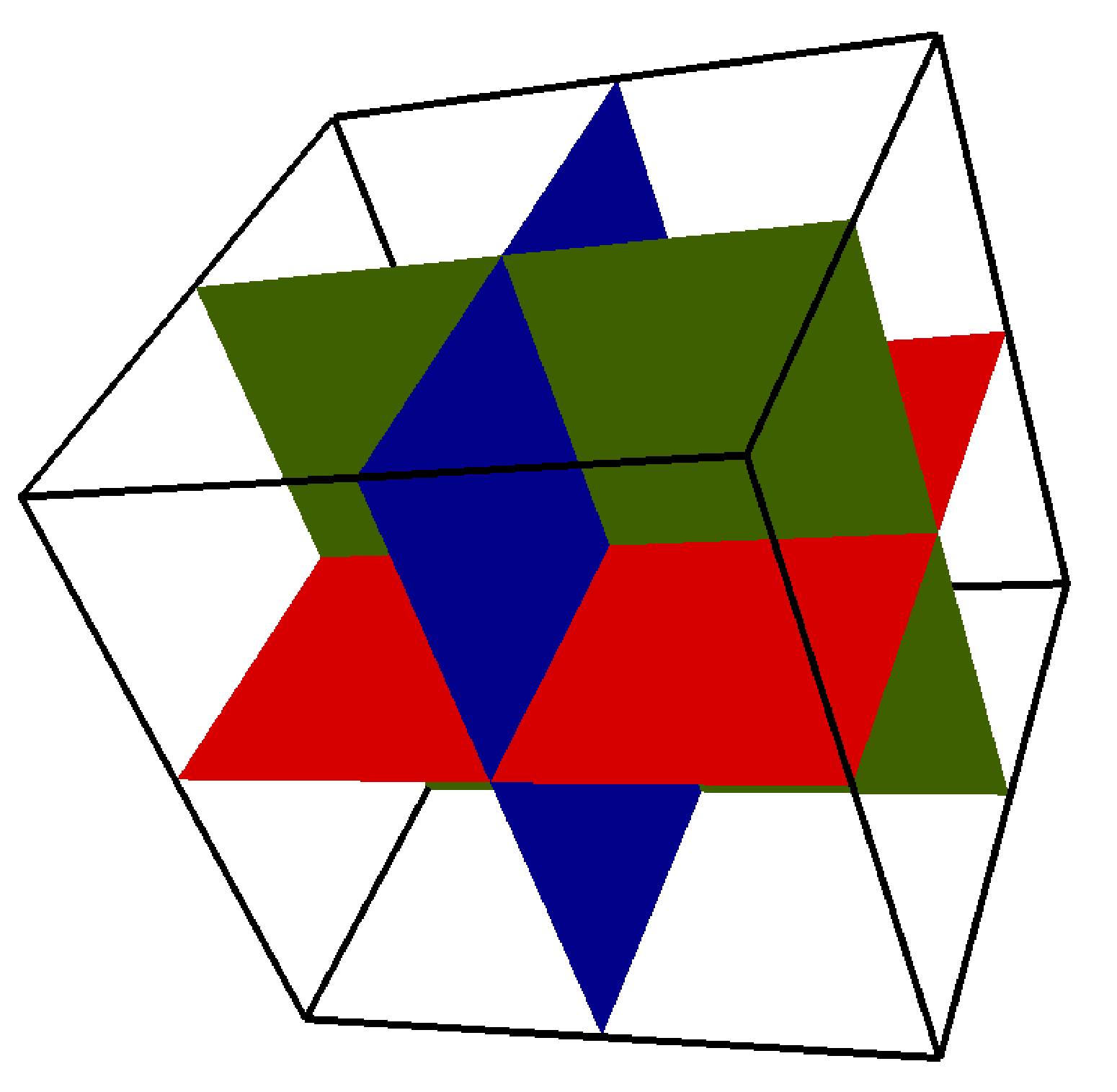}
\end{center}
\caption[]{
For a triply periodic box the three fundamental periods are periodic planes
through the domain with linearly independent surface normals.
Those three planes are highlighted in 3 different colors.
}
\label{fig: triple box}\end{figure}

Unlike the doubly periodic case, triply periodic domains are harder to
visualize as they need to be embedded in $\mathbb{R}^4$.
For the sake of simplicity we shall not attempt to visualize it in some
higher dimensional space and only outline the geometry of the fundamental
periods.
The fundamental periods are planes with normals equal to the unit vectors
(\Fig{fig: triple box}).

From the Poincar\'e polynomials we can calculate the Betti number \hl{of second kind} to $b_2 = 3$
with the Poincar\'e  polynomial
\EQ
P_{S\times S\times S} = 1 + 3z + 3z^2 + z^3.
\EN
Any magnetic field with a net flux through any of these planes will lead
to $\int_{L_i} \beta \ne 0$ which implies that a magnetic vector potential
cannot exist.
Hence, for a potential to exist, the mean magnetic field has to vanish.

\section{Magnetic Helicity}

Here we present cases of magnetic fields in periodic domains
for which we compute the magnetic helicity.
We use the case for which $\AAA$ does not exist and show
that already from a geometrical approach the magnetic helicity
is not well defined.
We also show cases for which $\AAA$ exists, but the magnetic helicity
\hl{
cannot be equated to the Gauss linking number.
However, this does not mean that its topological interpretation
is not valid.
}

\subsection{Doubly Periodic Domain}

Computing the magnetic helicity $\int_{V} \AAA\cdot\BB\ \dd V$ can be done if
the magnetic vector potential is given, or alternatively, it can be computed
as the linking of magnetic field lines using the Gauss linking number.
The latter can be applied even for periodic domains
\citep{Panagiotou-2015-300-533-JCompPhys}, however it does not allow
for a mean magnetic flux.
Here we illustrate difficulties arising from the computation of the magnetic
helicity for doubly periodic domains with a mean magnetic field
\hl{
through the non-periodic dimension
}
similar to \cite{Berger-1996-34-225-ApLC, Berger-1997-102-2156-JGeoPhRes,
Watson-Craig-2001-106-A8-JGeoRes}.

\begin{figure}[t!]\begin{center}
\includegraphics[width=0.5\columnwidth]{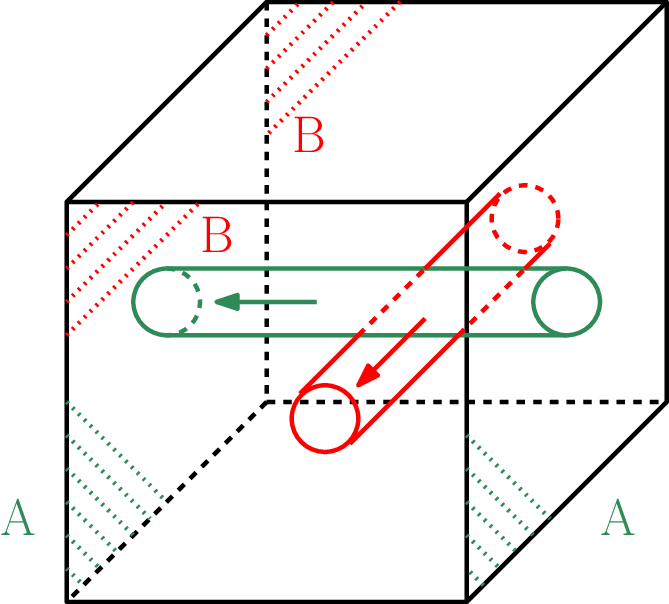} \\
\vspace{2ex}
\includegraphics[width=0.6\columnwidth]{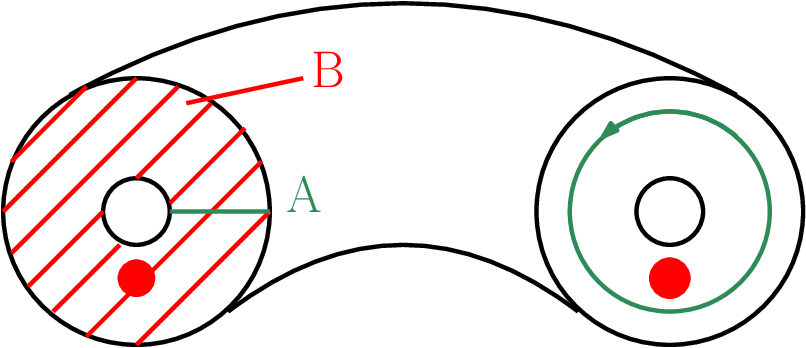}
\vspace{2ex}
\includegraphics[width=0.6\columnwidth]{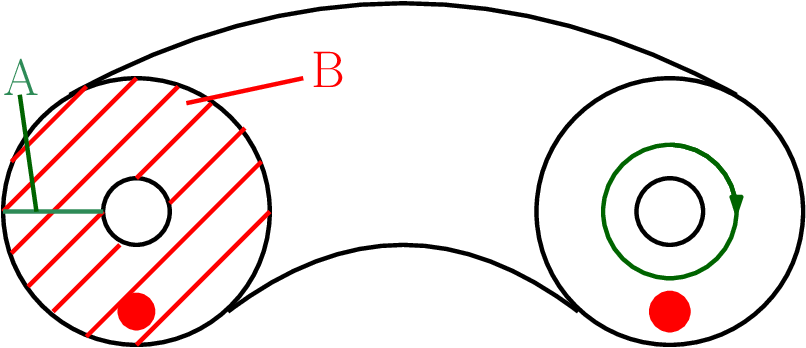}
\end{center}
\caption[]{
Two localized magnetic flux tubes with finite width in a cubic domain with two periodic
directions that align with the flux tubes (upper panel); represented as a toroidal
domain (central and lower panel).
In the cube representation, the green faces denoted as `A' are left and right,
while the red surfaces denoted as `B' are located at the front and back of the cube.
}
\label{fig: tubes_peri_peri}\end{figure}

Our test case consists of a doubly periodic box with
two localized finite width magnetic flux tubes
that both extend in the direction of the periodic dimension
(see \Fig{fig: tubes_peri_peri}, upper panel).
Since the doubly periodic domain is topologically equivalent to a solid
torus with a hole, we can represent this configuration as such
(see \Fig{fig: tubes_peri_peri}, central panel).
From the central picture we see that the two flux tubes are linked.
Using Gauss' integral we can therefore compute the linking and magnetic helicity
of this configuration, which will be non-zero.

However, there is an ambiguity in how we compose the toroidal domain.
We could have equally well joined together the sides `A' going the other way
(see \Fig{fig: tubes_peri_peri}, lower panel).
Using the Gauss' integral would lead to a vanishing magnetic helicity,
since in this realization, the flux tubes are not linked.
Rather than a physical effect,
\hl{
this is due to external linking caused by the choice of embedding
the periodic manifold in $\mathbb{R}^3$.
Unfortunately for the magnetic helicity this also means that we cannot
interpret this number in terms of the Gauss linking number.
}

\hl{
The geometric interpretation can nevertheless still be saved if we,
instead of using the Gauss linking number were to use the
flux averaged winding number of the magnetic field lines.
\cite{Prior-Yeates-2014-787-2-ApJ, Xiao-2025} showed that for periodic domains this quantity
is the magnetic helicity under the appropriate winding gauge.
}

\hl{
A transformation of the magnetic vector potential also has the power
to reduce the transformed helicity to zero.
To construct such an example and
}
for simplicity, imagine that the two flux tubes are localized such that their
magnetic fields do not overlap.
Furthermore, let us assume that their field is cylindrically symmetric around
their respective principal axis.
In such a geometry there is at least one gauge such that
the magnetic vector potentials $\AAA_1$ and $\AAA_2$ that generate
the fields in the tubes $\BB_1$ and $\BB_2$ are poloidal around the tubes, which implies
$\AAA_1 \cdot \BB_1 = \AAA_2 \cdot \BB_2 = 0$.

\hl{
Similar to a gauge transformation we add to the field $\AAA_1$ a closed,
but not necessarily exact, one-form $\GGGG_1$
\EQ
\AAA_1 \rightarrow \AAA_1 + \GGGG_1.
\EN
In a gauge transformation we would have a gradient of a periodic function here.
We take $\GGGG_1$ to be constant in the direction perpendicular to $\BB_2$
with constant component $\alpha_1$ along $\BB_2$.
With that we have
\EQ
\GGGG_1 \cdot \BB_2 = \alpha_1 |\BB_2|
\EN
Since $\nab\times\GGGG_1 = 0$ the transformation leaves the magnetic field unchanged.
But $\GGGG_1$ is not exact and would correspond to
\hl{
the gradient
}
\EQ
\GGGG_1 = \nab\Psi_1,
\EN
with the field
\EQ
\Psi_1 = \alpha_1 y,
\EN
which would not be single-valued in the periodic domain, assuming that
the axis of $\BB_2$ is $y$.
Instead, $\GGGG_1$ is the harmonic representative of the non-trivial
first cohomology class for the periodic $y$-direction.
Unlike a gauge transformation this gives us a closed non-exact
freedom of transformation.
}

With this the transformed magnetic helicity becomes
\EQ
H_{12}' = H_{12} + \int \GGGG_1\cdot\BB_2 \ \dd V.
\EN
The integral then modifies to
\EQ
H_{12}' = H_{12} + \alpha_1\phi_2 L_y,
\EN
with the magnetic flux $\phi_2$.
If we now choose the gradient $\alpha_1$ such that $\alpha_1 = -H_{12}/(\phi_2 L_y)$
the  transformed magnetic helicity $H'_{12}$ vanishes.

The same calculation can be done for $H_{21}$ with
\EQ
\AAA_2 \rightarrow \AAA_2 + \GGGG_2,
\EN
where
\EQ
\GGGG_2\cdot\BB_1 = \alpha_2 |\BB_1|
\EN
and
\EQ
\alpha_2 = -H_{21}/(\phi_1 L_x)
\EN
Here, the flux tube $1$ extends along the $x$-axis.

Of course we cannot choose two different transformations for $\AAA_1$
and $\AAA_2$, but we can combine the two such that the total helicity $H' = 0$.
First, note that $\GGGG_1\cdot\BB_1 = \GGGG_2\cdot\BB_2 = 0$ is a valid choice,
as it does not change our results on $H'_{12}$ and $H'_{21}$.
With that we can construct the transformation as
\EQ
\label{eq: full gauge transform}
\AAA \rightarrow \AAA + \GGGG_1 + \GGGG_2
\EN
with the same $\GGGG_i$ as before.
With that we have $H' = 0$.

For the doubly periodic domain with mean magnetic flux in the periodic
directions we showed that the magnetic vector potential $\AAA$ exists.
Despite its existence, the magnetic helicity is not well defined.
Changing the
\hl{
transformation fields,
}
we can even transform it away, which would not
make it physically meaningful.

\subsection{Unphysical Magnetic Flux Outside the Domain}

\hl{Even} before we apply our explicit transformation, \hl{an} implied gauge for the magnetic
vector potential generates a magnetic flux outside the domain which
can change the magnetic helicity content.
The applied \hl{transformation} does not change this.

To see this we first need to write down an expression for $\AAA_1$,
i.e.\ the flux tube in the $x$-direction.
To generate a magnetic field in the $x$-direction we can find
\begin{equation}
\AAA_1 =
\left\{
\begin{array}{l}
|\BB_1|/2 (y\ee_z - z\ee_y), \qquad \mathrm{inside\ the\ tube} \\
|\BB_1|r_0^2 /2 \frac{y\ee_z - z\ee_y}{y^2 + z^2} , \qquad \mathrm{outside\ the\ tube}
\end{array}
\right.,
\end{equation}
where $r_0$ is the radius of the tube.

\begin{figure}[t!]\begin{center}
\includegraphics[width=0.5\columnwidth]{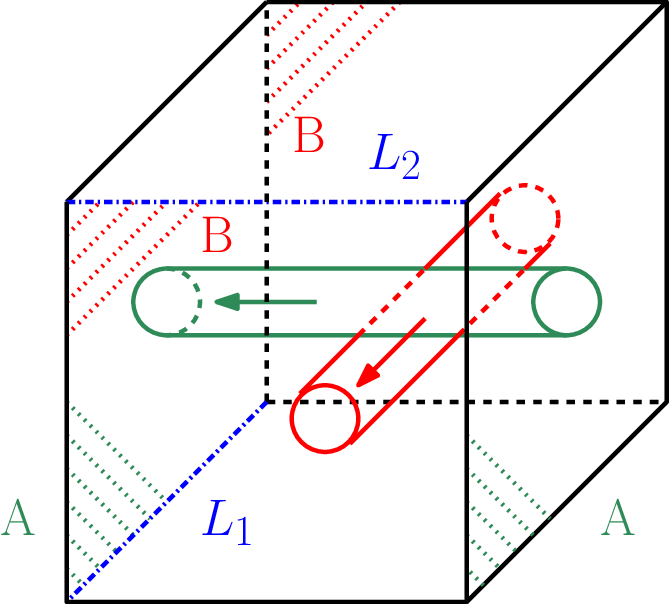} \\
\vspace{2ex}
\includegraphics[width=0.5\columnwidth]{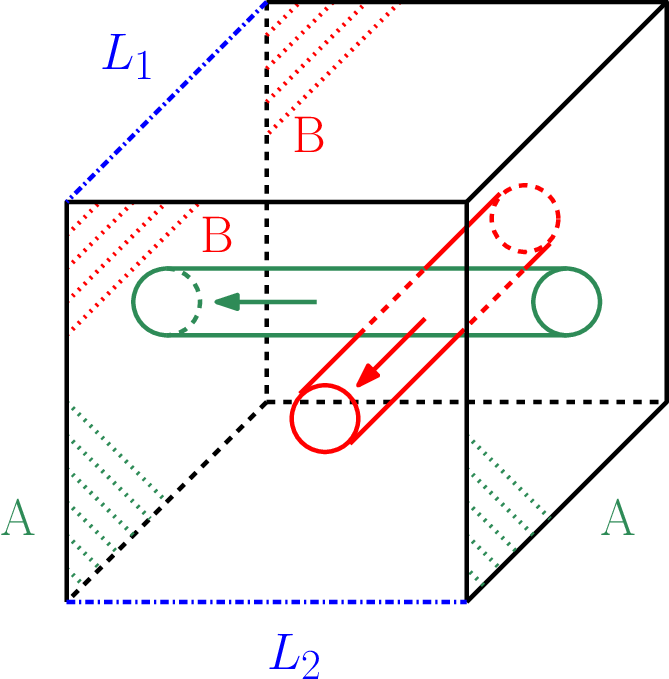} \\
\vspace{2ex}
\includegraphics[width=0.6\columnwidth]{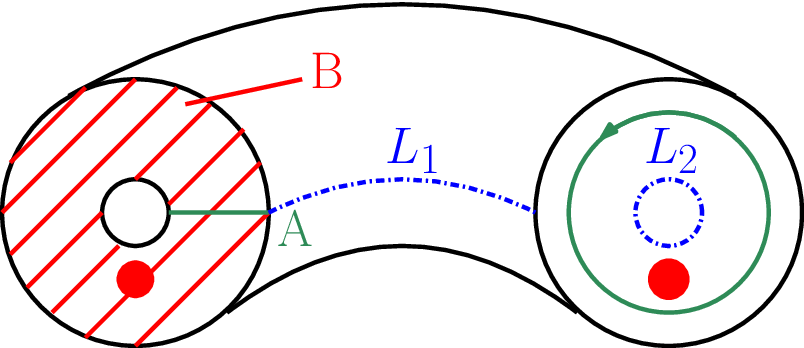}
\vspace{2ex}
\includegraphics[width=0.6\columnwidth]{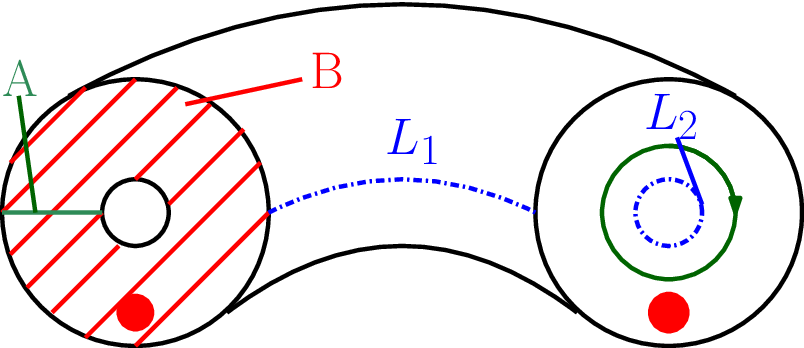}
\end{center}
\caption[]{
The same localized magnetic flux tubes as in \Fig{fig: tubes_peri_peri}
but now with the integration domains $L_1$ and $L_2$ for determining the
unphysical magnetic flux outside the domain.
}
\label{fig: tubes_peri_peri_cube_linton}\end{figure}

We now show that there is a net magnetic field outside the domain by
calculating the magnetic flux through the $L_1$ ring (see \Fig{fig: tubes_peri_peri_cube_linton}), which lies on the domain boundary.
This is the loop integral
\EQ
\phi_1^\mathrm{out} = \oint_{L_1} \AAA\cdot\tt \ \dd \llll.
\EN
To make our calculations easier, we set the box corners to $[-1, -1, -1]$
and $[1, 1, 1]$.
With our transformation we then have
\EQA
\phi_1^\mathrm{out} & = & \oint_{L_1} (\AAA_1 + \GGGG_1 + \GGGG_2) \cdot\tt \ \dd \llll \nonumber \\
 & = & \oint_{L_1} (\AAA_1 + \GGGG_1 + \GGGG_2) \cdot\BB_2/|\BB_2| \ \dd l \nonumber \\
 & = & \oint_{L_1} (\AAA_1 + \GGGG_1) \cdot\BB_2/|\BB_2| \ \dd l \nonumber \\
 & = & \oint_{L_1} \left(\frac{|\BB_1| r_0^2}{2}\frac{1}{1 + l^2} + \alpha_1\right)\ \dd l \nonumber \\
 & = & |\BB_1|r_0^2 \pi/4 + L_y \alpha_1 \nonumber \\
 & \ne & 0.
\ENA
Similarly we can show that in this transformation $\phi_2^\mathrm{out} \ne 0$.

If these integrals are non-zero we have a magnetic flux through the loops
$L_1$ and $L_2$ that lies {\em outside} the domain.
Since this flux is linked with the inner magnetic field it adds
to the magnetic helicity.

\subsection{Magnetic Helicity in Multiply Connected Domains}

Any of the shown domains with at least one periodic dimension
is a multiply connected domain.
For such domains, it has been suggested to use an alternative
definition of the magnetic helicity that is invariant under gauge
transformations \citep{Mactaggart-Valli-JPlasmPh-85-5-2019}.
For our toroidal shell its general form reduces to
\EQA
\Upsilon(\BB) & = &
\int \AAA\cdot\BB\ \dd V -
\oint_{\gamma_1}\AAA\cdot\BoldVec{t}_1 \int_{S_1}\BB\cdot\nn_1 - \nonumber \\
 & & \oint_{\gamma_2}\AAA\cdot\BoldVec{t}_2 \int_{S_2}\BB\cdot\nn_2,
\ENA
where $\gamma_i$ are closed paths that connect across the periodic dimensions,
$\BoldVec{t}_i$ the tangents to $\gamma_i$, $S_i$ the surface area
of the connecting surfaces and $n_i$ the normals to the surfaces.
In our simple thought experiment $\gamma_i$ are straight lines along $\BB_i$.
With that we have $\BoldVec{t}_i = \BB_i/|\BB_i|$.

We can now compute the transformed $\Upsilon'(\BB)$ using
$\AAA' = \AAA + \GGGG_1 + \GGGG_2$ from equation \eqref{eq: full gauge transform} as
\EQ
\Upsilon'(\BB) =
\int\AAA'\cdot\BB\ \dd V
- \oint_{\gamma_1}\AAA'\cdot\BoldVec{t}_1\phi_1
- \oint_{\gamma_2}\AAA'\cdot\BoldVec{t}_2\phi_2,
\EN
where we used $\int_{S_i}\BB\cdot\nn_i = \phi_i$.
We know that with our transformation the helicity vanishes, i.e.\
$\int\AAA'\cdot\BB\ \dd V = 0$.
So with $\GGGG_i\cdot\BB_i = 0$ our transformed helicity simplifies to
\EQA
\Upsilon'(\BB) & = &
-\oint_{\gamma_1}\AAA\cdot\BoldVec{t}_1\phi_1 - \oint_{\gamma_2}\AAA\cdot\BoldVec{t}_2\phi_2 \nonumber \\
 & & - \oint_{\gamma_1}\GGGG_2\cdot\BB_1/|\BB_1| \nonumber \\
 & & - \oint_{\gamma_2}\GGGG_1\cdot\BB_2/|\BB_2|.
\ENA
Now we use our definitions for $\GGGG_i$ from previously and obtain
\EQA
\Upsilon'(\BB) & = &
-\oint_{\gamma_1}\AAA\cdot\BoldVec{t}_1\phi_1 - \oint_{\gamma_2}\AAA\cdot\BoldVec{t}_2\phi_2 \nonumber \\
 & & -\oint_{\gamma_1}\alpha_2|\BB_1|/|\BB_1|\phi_1 \nonumber \\
 & & - \oint_{\gamma_2}\alpha_1|\BB_2|/|\BB_2|\phi_2.
\ENA
We defined $\alpha_i$ in order to obtain $H' = H'_{12} = H'_{21} = 0$.
With their definition we have
\EQA
\Upsilon'(\BB) & = &
-\oint_{\gamma_1}\AAA\cdot\BoldVec{t}_1\phi_1 - \oint_{\gamma_2}\AAA\cdot\BoldVec{t}_2\phi_2 \nonumber \\
 & & +\oint_{\gamma_1}H_{21}/(\phi_1 L_y)\phi_1 \nonumber \\
 & & + \oint_{\gamma_2}H_{12}/(\phi_2 L_z)\phi_2,
\ENA
and finally
\EQA
\Upsilon'(\BB) & = &
-\oint_{\gamma_1}\AAA\cdot\BoldVec{t}_1\phi_1 - \oint_{\gamma_2}\AAA\cdot\BoldVec{t}_2\phi_2 \nonumber \\
 & & + H_{12} + H_{21} \\
 & = & \Upsilon(\BB).
\ENA
Hence, the helicity definition for multiply connected domains is invariant
for our example using our transformation that would lead to zero helicity.

\subsection{\hl{Mean} Flux Through the non-Periodic Dimension}

\begin{figure}[t!]\begin{center}
\includegraphics[width=0.5\columnwidth]{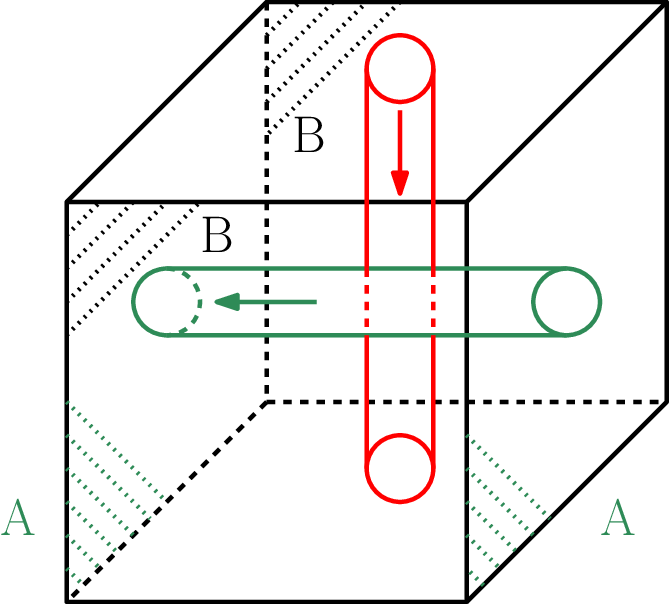} \\
\vspace{2ex}
\includegraphics[width=0.6\columnwidth]{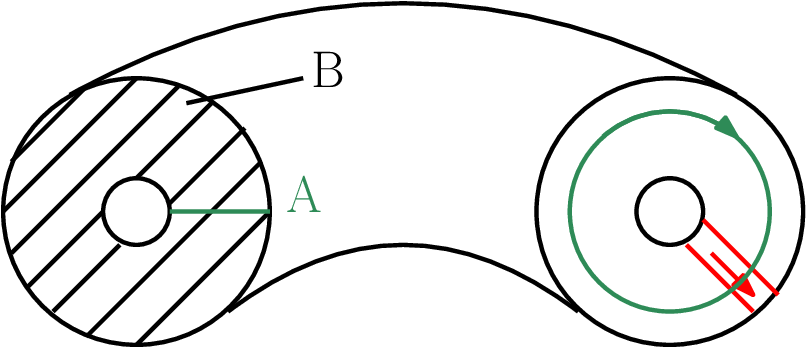}
\end{center}
\caption[]{
Two localized magnetic flux tubes with finite width in a cubic domain with two periodic
directions one of which aligns with one of the flux tubes (upper panel); represented as a toroidal
domain (lower panel).
}
\label{fig: tubes_peri}\end{figure}

We now change our set up by orienting one of the flux tubes in the direction of the
non-periodic dimension (see \Fig{fig: tubes_peri}, upper panel).
If we represent this domain as a solid torus with a hole embedded in $\mathbb{R}^3$
we see that the two flux tubes are not linked.
Furthermore, the tube along the non-periodic direction is not closed
which means that we cannot compute any Gauss integral and cannot compute
any magnetic helicity.
In this counter example a magnetic field is constructed for which no vector potential exists,
as argued further above.
However, we could argue without such knowledge, on the non-existence of
magnetic helicity for such configurations.

\subsection{Zero Mean Flux Through the non-Periodic Dimension}
\hl{
All of our considerations about the existence of a magnetic vector potential
and the physical interpretation of the magnetic helicity distinguished
cases with a mean magnetic flux through one of the dimensions or without.
For solar magnetic fields these cases are certainly important to distinguish.
However, we would like to point out a finer distinction between
zero flux at the boundary, as discussed in some of our examples,
and a vanishing surface integral of the magnetic field's normal component.
Closed magnetic flux tubes that are entirely located within the domain
would lead to zero flux at the boundary.
Such configurations are of theoretical importance.
For solar magnetic fields, however, we have to consider the case
of non-zero flux.
}

\hl{
For solar magnetic field research we often limit ourselves
to subdomains of the Sun's atmosphere.
These are typically cuboids for which we
know the magnetic field at their lower bound, the photosphere, from observations.
Magnetic field extractions can then give us a proxy for the
field in the full box, including all its boundaries.
While the mean magnetic flux through the boundaries is zero,
we are not always dealing with periodic or perfect conductor boundaries.
This has no bearing on the existence of the magnetic vector potential,
however we end up with a magnetic helicity that is not gauge-independent.
}

\hl{
For a gauge transformed magnetic vector potential we have
\EQ
\AAA' = \AAA + \nab\phi,
\EN
with the differentiable scalar field $\phi$.
The magnetic helicity transforms as
\EQA
H' & = & \int_V \AAA'\cdot\BB \ \dd V \nonumber \\
 & = & \int_V (\AAA + \nab\phi)\cdot\BB \ \dd V \nonumber \\
 & = & H + \int_V \nab\phi\cdot\BB \ \dd V.
\ENA
The second term can be transformed
into a surface integral using Gauss's theorem
\EQ\label{eq: helicity gauge}
\int_V \nab\phi\cdot\BB\ \dd V = \int_{\partial V}\phi\BB\cdot\nn \ \dd S.
\EN
For periodic or perfect conductor boundaries this term vanishes.
But if we have a non-vanishing magnetic field through a non-periodic
boundary, even with vanishing mean flux, this term is in general non-zero.
}

\hl{
However, we can make it zero with a non-trivial gauge.
As example we use a system that is non-periodic in the $z$-direction.
For the side boundaries we force periodicity while the top
and bottom boundaries are non-periodic but also not perfectly conducting.
With that the top and bottom magnetic field is allowed to differ from zero
while the mean flux vanishes.
Let's construct the field at the two vertical boundaries such that
it is localized in two circles with opposite polarity.
With symmetry around the $y$-axis and equal magnitude
the surface integrated magnetic flux is zero on each vertical boundary,
as required by the divergence freeness.
To make it non-periodic we can choose $\BB$ at each boundary
to differ only by a scalar $\alpha \ne 1$:
\EQ\label{eq: bottom top bb}
\BB(x, y, z_\mathrm{bottom}) = \alpha \BB(x, y, z_\mathrm{top}).
\EN
This is a simplified case for the solar application where
the total flux through the boundaries is zero, while the field
is neither zero nor periodic.
}

\hl{
We can now apply the gauge on our gauge transformed helicity
\eqref{eq: helicity gauge}:
\EQ
H' = H + \int_\mathrm{bottom}\phi\BB\cdot\nn \ \dd S + \int_\mathrm{top}\phi\BB\cdot\nn \ \dd S,
\EN
with the gauge field $\phi$.
Note that unlike for the periodic case, here $\phi$ does not need
to be chosen to be periodic in $z$.
Multiplying the magnetic field with the surface outwards pointing normal vector
we can then write
\EQA
H' & = & H - \int\phi(x, y, z_\mathrm{bottom})B_z(x, y, z_\mathrm{bottom})\ \dd x \ \dd y + \nonumber \\
 & & \int\phi(x, y, z_\mathrm{top})B_z(x, y, z_\mathrm{top})\ \dd x \ \dd y.
\ENA
Using the proportional relation \eqref{eq: bottom top bb} we can write
\EQA
H' & = & H - \alpha\int\phi(x, y, z_\mathrm{bottom})B_z(x, y, z_\mathrm{top})\ \dd x \ \dd y + \nonumber \\
 & & \int\phi(x, y, z_\mathrm{top})B_z(x, y, z_\mathrm{top})\ \dd x \ \dd y \\
 & = & H + \int\left(\phi(x, y, z_\mathrm{top}) - \alpha\phi(x, y, z_\mathrm{bottom})\right)
 B_z(x, y, z_\mathrm{top})\ \dd x \ \dd y.
\ENA
If we now choose for the gauge field $\phi$ the form
\EQ
\phi(x, y, z) = \beta B_z(x, y, z_\mathrm{top}),
\EN
with $\beta \ne 0$ we get
\EQ
H' = H + \int(1-\alpha)\beta B_z^2(x, y, z_\mathrm{top})\ \dd x \ \dd y.
\EN
If we choose
\EQ
\beta = \frac{-H}{(1 - \alpha)\int B_z^2(x, y, z_\mathrm{top})\ \dd x \ \dd y}
\EN
our gauge field becomes
\EQ
\phi(x, y, z) = \frac{-H B_z(x, y, z_\mathrm{top})}{(1 - \alpha)\int B_z^2(x, y, z_\mathrm{top})\ \dd x \ \dd y}
\EN
we obtain $H' = 0$.
We should stress that this condition on $\phi$
only needs to be imposed at the top and bottom boundaries.
Internally we could still have a freedom of choice as long as it is
consistent with the top and bottom boundary and differentiable.
}

\hl{
We can now choose the two foot points and their induced flux tubes
to have an internal twist.
The normal component of $\BB$ is not affected by that.
This means that we can still find a gauge for which $H' = 0$
even with a net magnetic twist.
Such twist would be picked up by using the winding gauge
\citep{Prior-Yeates-2014-787-2-ApJ, Xiao-2025} which would return $0$ in the absence of twist
and a non-zero value with twist, as the helicity
is uniquely determined by $\BB$ there.
So, in this simple example the winding gauge helicity is non-zero,
while we can still find a gauge transformation for which $H' = 0$
which carries no geometrical interpretation.
For a solar relevant open flux example the winding helicity
retains its geometrical interpretation.
}

\subsection{Triply Periodic Domain: Numerical Examples}

\begin{figure}[t!]\begin{center}
\includegraphics[width=0.4\columnwidth]{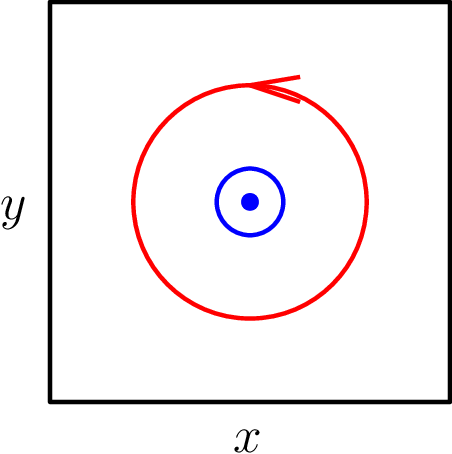} \qquad
\includegraphics[width=0.4\columnwidth]{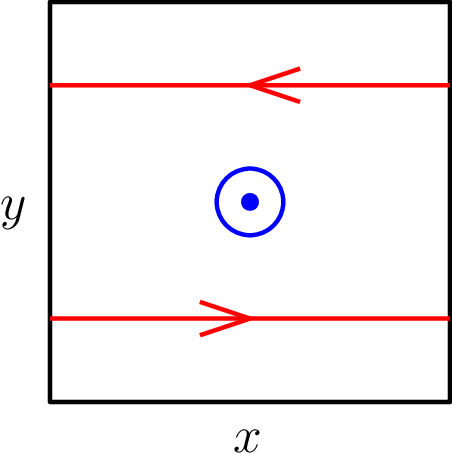}
\end{center}
\caption[]{
Schematic representation of the magnetic flux tubes in a $xy$-plane projection
used to compute the magnetic helicity content in a triply periodic domain.
We obtain the configuration on the right via topology conserving transformations
of the left configuration
through reconnecting the flux ring with itself through the $x$-boundaries.
}
\label{fig: tube_ring}\end{figure}

To demonstrate the contradictions of magnetic helicity in periodic domains further,
we perform numerical calculations of the magnetic helicity for a field with mean flux.
It consists of a flux tube in the $z$-direction of the form
\EQ
\BB_{\rm vertical} = \exp{(-8(x^2+y^2))} \ee_z,
\EN
and a ring of the form
\EQA
\BB_{\rm ring} & = & \exp{(8(-(\sqrt{x^2+y^2}-2\pi/3)^2-z^2))} \nonumber \\
 & & \times (-y\ee_x + x\ee_y)/\sqrt{x^2 + y^2}.
\ENA
For both components we chose the profile to be Gaussian so that
we can assume the tubes to be localized to a high degree.
The whole magnetic configuration is then simply
\EQ
\BB = \BB_{\rm vertical} + \BB_{\rm ring}.
\EN
A schematic representation is shown as projection on the $xy$-plane in \Fig{fig: tube_ring}
(left panel).
We then deform the field
\hl{in a topology conserving way}
by extending and
reconnecting it through the $x$-boundaries such that the initial ring becomes
two anti-parallel flux tubes with the same cross section (\Fig{fig: tube_ring}, right panel).

Both test cases are placed in a cubic box of size $2\pi$ with $256$ grid points.
We then compute a vector potential $\AAA$ for these fields using the Fourier transforms,
since we know that
\EQA
\JJ & = & \nab\times\BB \nonumber \\
 & = & \nab\times\nab\times\AAA \nonumber \\
 & = & -\nab^2\AAA,
\ENA
with the Coulomb gauge $\nab\cdot\AAA = 0$.
In Fourier space this simplifies to the algebraic equation
\EQ
\hat{\JJ} = k^2\hat{\AAA},
\EN
with the Fourier transformed current density $\hat{\JJ}$ and magnetic vector potential $\hat{\AAA}$.
We can use this method, since the fields are periodic.
\hl{
In a final step we perform the inverse Fourier transform to obtain
the magnetic vector potential in real space
\EQ
\AAA = \mathcal{F}^{-1}\{\hat{\JJ}/k^2\}.
\EN
}

With $\AAA$ and $\BB$ we then compute the magnetic helicity in our computational
domain and obtain different values for the two cases.
Case one with the ring results in a magnetic helicity of $H_{\rm m} \approx 0.2$
\hl{in code units},
while the second configuration results in $H_{\rm m} \approx 0.1$.
It is clear that a topology conserving transformation should not have resulted in
different numbers.
Furthermore, an additional deformation through the $y$-boundaries of the two anti-parallel tubes
would result in a flux ring that does not enclose the vertical flux tube.
This configuration would not be helical.

Similar calculations were performed by \cite{Linton-Antiochos-2005-625-506-ApJ} for a doubly
periodic domain.
One configuration consisted of two magnetic flux tubes that cross at an orthogonal
angle (see \Fig{fig: tubes_peri}).
They argued that such a configuration should have an integer linking number of $2$,
where the linking number is simply $L = H_{\rm m}/\phi^2$, with the axial
magnetic flux $\phi$.
They applied a gauge transformation to eliminate any unphysical net magnetic flux
outside the domain.
For us their gauge would translate to
\EQA
\AAA' & = & \AAA - \frac{\ee_x}{x_1 - x_0} \int_{x_0}^{x_1} A_x(x, y_0, z_0)\ \dd x \nonumber \\
 & & - \frac{\ee_z}{z_1 - z_0} \int_{z_0}^{z_1} A_z(x_0, y_0, z)\ \dd z,
\ENA
where $(x_0, y_0, z_0)$ is the origin of the coordinate system.
In our calculations we do not find an integer for $L$, instead for us
$L \approx 0.679249$.

Similarly, they predicted that for a diagonal flux tube in such a multiply connected domain
the magnetic helicity should be finite, i.e.\ $\pm 1$.
However, in our calculations we obtain $0$.
These two examples by \cite{Linton-Antiochos-2005-625-506-ApJ} show once again the
ambiguity of computing the magnetic helicity in such domains
where an unambiguous topological interpretation is not possible.
We therefore suggest that such calculations should be done with a critical mind.

\subsection{Flux-Freeness as Restriction for Reconnection}

In the previous two sections we demonstrated that by rearranging and reconnecting
the magnetic field we can turn a seemingly helical field into a non-helical field.
Here we demonstrate with a simple example, the Hopf link, how the absence of a net
magnetic flux in the triply periodic case restricts the freedoms on how we can
reconnect the field.
This preserves the meaningfulness of helicity in triply periodic domains.

\begin{figure}[t!]\begin{center}
\includegraphics[width=0.4\columnwidth]{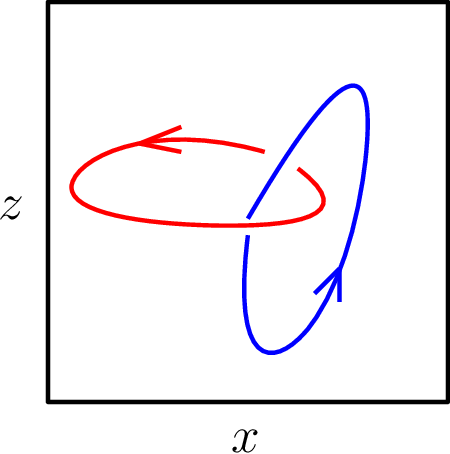} \qquad
\includegraphics[width=0.4\columnwidth]{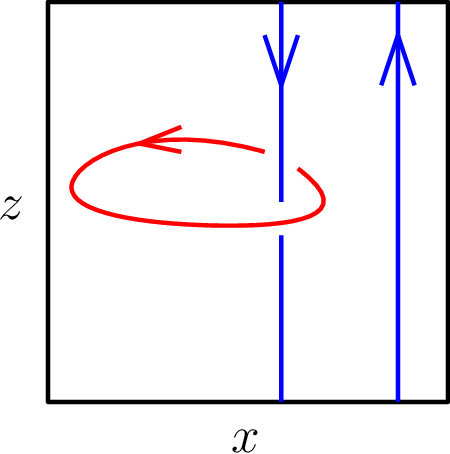} \\
\vspace{4ex}
\includegraphics[width=0.4\columnwidth]{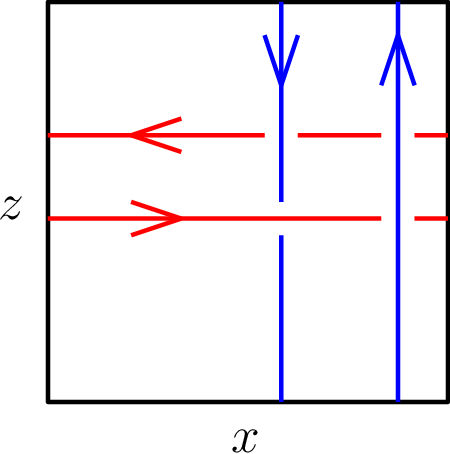} \qquad
\includegraphics[width=0.4\columnwidth]{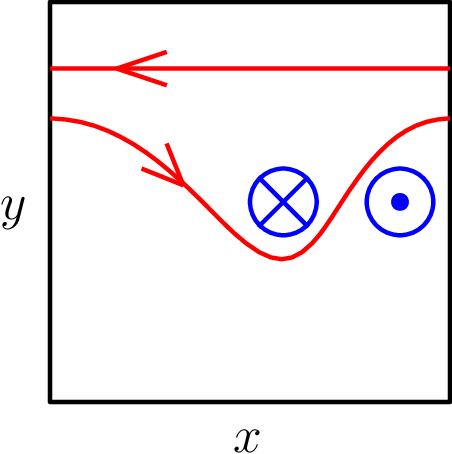}
\end{center}
\caption[]{
Schematic representation of the magnetic Hopf link (upper left).
Deformations and reconnection of the two flux rings through the $z$ (upper right)
and $x$ boundaries (lower left) straighten out the rings.
Seen from top ($xy$-projection) we see that one of the red flux tubes must wind
around the blue (vertical) tubes, conserving magnetic helicity.
}
\label{fig: hopf}\end{figure}

Consider the Hopf link in a triply periodic domain (\Fig{fig: hopf}, upper left).
This configuration is helical.
We can deform the flux ring on the right such that it reconnects through the boundaries
(\Fig{fig: hopf}, upper right).
The same can be done with the flux ring to the left (\Fig{fig: hopf}, lower left).
If we were to observe this configuration from the top, i.e.\ in a projection in the
$xy$-plane, we would see the red flux tube winding around the two straight blue flux tubes
(\Fig{fig: hopf}, lower right).

We can now reconnect the two red flux tubes such that they form one ring encircling the left blue
flux tube in the anti-clockwise sense.
We can also reconnect the winding flux tube around the blue flux tube to the right in a clockwise
sense.
Both lead to the same sign of magnetic helicity.
It is also clear that the red flux tube cannot be reconnected such that it does not encircle any of the
blue flux tubes, preserving the magnetic helicity content.
Note that by removing one of the blue flux tubes one could perform such a reconnection operation.
However, that field would contain a net magnetic flux.

\section{Discussion}

For singly periodic domains we showed that any magnetic field distribution
can be generated by either internal or external (outside the domain)
currents.
For doubly periodic domains this can be achieved as well, as long as any mean
magnetic field does not point along the non-periodic direction.
If it points along the non-periodic direction then such a field is
generated by an un-physical magnetic monopole distribution.
In case of a triply periodic domain, any mean magnetic field would
need to be generated by an external current.
However, if we think of such a domain as a repetition of stacked boxes
we see that the field fills the entire space $\mathbb{R}^3$ which does
not leave any room for the generating currents.

With the help of differential forms and potentials we have pointed out the
non-existence of the magnetic vector potentials for doubly
\hl{
periodic domains and a mean flux through the non-periodic dimension,
and triply periodic domains with any mean flux,
using a more accessible description than \cite{Pfefferle-Noakes-2021-62-9-JMthPhys}.
}
Such combinations of magnetic fields and periodic domains while using
the magnetic vector potential do sometimes occur
\citep{KleeorinRuzmaikin82, BlackmanBrandenb2002ApJ,
BrandenbSubramanianReview2005, DelSordoTurbulence12,
GrahamBlackmanMinuteHelicity12, alpha2_periodic13,
Park-2013-434-3-MNRAS, Mitra-Brandenburg-2014-445-761-ApJ,
Vishniac-Shapovalov-2014-780-2-ApJ, Jabbari-Brandenburg-2015-805-2-ApJ,
Losada-Brandenburg-2014-564-A2-AA}
and care has to be taken in such situations.
Applications where $\AAA$ is used in periodic domains are dynamo generated
fields in turbulent plasma and magnetic flux tubes embedded in a box.
Both fields, however, do not have a mean flux through any of the sides.
The mean magnetic fields from dynamo calculations which arise from the
inverse cascade have positive and negative contributions at the boundaries
which then have a zero net contribution to the total flux.

For non-periodic domains the two-form $\beta$ is exact and a potential
$\alpha$ always exists.
Knowing $\BB$ one can then, in principle, compute $\AAA$.
Some of the methods, however, make use of the Fourier transformed fields
which \hl{implies} periodicity.
For a homogeneous field of the form $\BB(\xx) = B_0 \ee_z$ the resulting
vector potential would then be potentially $0$.
To circumvent this practical issue one can resort to other methods, e.g.
Volterra's formula.

For cases for which the magnetic vector potential is not defined,
the magnetic helicity density $\AAA\cdot\BB$ and total magnetic helicity
$\int_V \AAA\cdot\BB$ cannot be meaningfully defined either.
This was already pointed out by
\cite{Berger-1996-34-225-ApLC, Berger-1997-102-2156-JGeoPhRes}
who showed in a thought experiment that magnetic helicity can be reversed
in sign in a doubly periodic box with a mean flux simply through permissible
reconnection events.
This then led to numerical simulations in two dimensions of such a field
by \cite{Watson-Craig-2001-106-A8-JGeoRes} who confirmed such a
helicity reversal.
Here we showed an example for a doubly periodic domain for which the magnetic
helicity cannot be defined without making use of the vector potential.

\hl{
Such constructions show that a straight forward interpretation
of the magnetic helicity as the
Gauss linking number is not always possible.
Using the winding gauge to compute the winding helicity fixes this
for such cases \citep{Prior-Yeates-2014-787-2-ApJ, Xiao-2025}.
We then have to equate the helicity with the flux averaged magnetic field
winding, which gives us the geometric interpretation.
}

\hl{
For solar relevant magnetic fields with non-periodic boundaries
and pointwise non-zero magnetic field through the boundaries we showed
that an appropriate gauge transformation can lead to
zero magnetic helicity.
The winding gauge correctly tracks this, returning zero in case of
no twist and non-zero with twist.
}

\hl{
If a magnetic vector potential exists, it still leaves the gauge
freedom, which makes the field non-unique.
Any closed, but not necessarily exact, one-form can be added
which would result in the same magnetic field when the curl is applied.
The implications on the magnetic helicity are evident.
Conversely, however, we cannot deduce anything about the existence
simply from the failure of a gauge choice or the Gauss
linking number to give us a meaningful helicity.
}

Authors like \cite{Matthaeus-Goldstein-1982-87-A8-JGeoRes, Brandenburg-Matthaeus-2004-69-7-PRE,
Stribling-Matthaeus-1995-2-1437-PhysPlasm}
already recognized that in presence of a mean magnetic field, the magnetic helicity is not
conserved under an ideal evolution.
This was circumvented by adding a component that involves the mean magnetic field.
However, it loses its interpretation as a measure for the field line topology.

\begin{acks}[Acknowledgments]
The author is grateful for the fruitful discussions with
Gunnar Horning at the University of Dundee and
David MacTaggart at the University of Glasgow.
\hl{
All comments by the anonymous referee have been gratefully received,
which has led to a significant improvement of the paper.
}
This work has been made with the financial support by the
Deutsche Forschungsgemeinschaft
(DFG, German Research Foundation) through the research unit FOR 5409
"Structure-Preserving Numerical Methods for Bulk- and Interface Coupling of Heterogeneous Models (SNuBIC)" (project number 463312734).
\end{acks}

\bibliographystyle{spr-mp-sola}
\bibliography{references}

\end{document}